# Organic solvent boosts charge storage and charging dynamics of conductive MOF supercapacitors

*Ming Chen, Taizheng Wu, Liang Niu, Ting Ye, Wenlei Dai, Liang Zeng, Alexei A. Kornyshev, Zhenxiang Wang, Zhou Liu, Guang Feng[*]*

M.C., T.Z.W., L.N., T.Y., W.L.D., L.Z., Z.X.W., Z.L., G.F.

State Key Laboratory of Coal Combustion, School of Energy and Power Engineering, Huazhong University of Science and Technology (HUST), Wuhan 430074, China.

A.A.K.

Department of Chemistry, Faculty of Natural Sciences, Imperial College London, Molecular Sciences Research Hub, White City Campus, W12 0BZ, London, United Kingdom

Email: gfeng@hust.edu.cn

## Abstract:

Conductive metal-organic frameworks (c-MOFs) and ionic liquids (ILs) have emerged as auspicious combinations for high-performance supercapacitors. However, the nanoconfinement from c-MOFs and high viscosity of ILs slow down the charging process. This hindrance can, however, be resolved by adding solvent. Here, we performed constant-potential molecular simulations to scrutinize the solvent impact on charge storage and charging dynamics of MOF-IL-based supercapacitors. We find conditions for >100% enhancement in capacity and ~6 times increase in charging speed. These improvements were confirmed by synthesizing near-ideal c-MOFs and developing multiscale models linking molecular simulations to electrochemical measurements. Fundamentally, our findings elucidate that the solvent acts as an 'ionophobic agent' to induce a substantial enhancement in charge storage, and as an 'ion traffic police' to eliminate convoluted counterion and co-ion motion paths and create two distinct ion transport highways to accelerate charging dynamics. This work paves the way for the optimal design of MOF supercapacitors.

## Keywords:

# 1. Introduction

Supercapacitors, recognized for their high power density, rapid charge-discharge capability, and excellent cycle life, have become compelling candidates for next-generation electrical energy storage.[1] Since the energy is stored in the electrode-electrolyte interface via the formation of electric double layers (EDLs), the straightforward and effective approach to enhancing the capacitive performance of the supercapacitor significantly relies on the development of porous electrode materials and advanced electrolytes.[1-2] Owing to the high electrical conductivity, controllable pore sizes, and monodispersed pores, conductive metal-organic frameworks (c-MOFs) have emerged as novel and designable electrode materials for supercapacitors.[3] Ionic liquids (ILs), featured with non-volatility, outstanding thermal stability, and wide electrochemical window, have been taken as promising electrolytes.[4] In particular, recent efforts in modeling and experimentation have highlighted the potential of combining c-MOFs and ILs to improve the energy storage performance.[5]

Despite these advancements, the development of MOF-IL-based supercapacitors remains a significant challenge. Experiments and molecular simulations have illuminated that the nanoconfinement from nanoporous electrodes and the high viscosity of ILs can hinder ion transport, slowing down the charging dynamics.[6] Molecular dynamics (MD) simulations with nanoslits have uncovered that a proper matching of the nanopore size and ion dimension holds the potential to simultaneously enlarge the capacitance and power density, owing to the 3D structure translation of in-pore ion packing.[7] However, achieving such matching is impeded by the complex ion interactions. Prior experimental and theoretical studies in porous carbon supercapacitors have demonstrated that adding organic solvents could enhance power density.[8] However, the enhancement in capacitance by adding solvents is limited, up to 50% for porous electrodes with a 1.3 times increase in charging process;[9] the capacitance even deteriorated in some porous carbon materials, owing to various topologies and connections of porous electrodes.[10] Meanwhile, the impact of solvent on the performance of porous carbon supercapacitors has been extensively studied, while the underlying polarization phenomena and energy storage mechanism of solvent effects on MOF-IL supercapacitors are in their infancy.[5b, 11] Therefore, the pursuit of expedited charging speed with enlarged capacity still faces a formidable obstacle.

Herein, to elucidate the effect of solvent on energy storage of supercapacitors based on c-MOF electrodes and IL electrolytes, we conducted constant-potential MD simulations with $Ni_3$(2,3,6,7,10,11-hexaiminotriphenylene)$_2$ ($Ni_3(HITP)_2$) electrode incorporated in 1-butyl-3-methylimidazolium hexafluorophosphate ([Bmim][PF$_6$]) with adding acetonitrile (ACN). Details of simulations are presented in **Methods**. Results reveal that the addition of ACN with an optimal ratio of IL to solvent (0.107) can significantly boost the charge storage capacity and charging speed *simultaneously*. To confirm these enhancements, we promoted the synthesis of $Ni_3(HITP)_2$ to obtain a specific surface area (SSA) close to its theoretical value of MOF crystal, and further developed multiscale models for cyclic voltammogram (CV) and galvanostatic charge and discharge (GCD), linking microscopic properties to macroscopic performance. Essentially, the enhancements in charge storage and charging dynamics could be attributed to the induced ionophobicity of electrolytes[4a, 12] and the construction of two distinct ion transport highways in c-MOF pores, respectively.

## 2. Results and discussion

**2.1 Effects of adding solvent on capacitive performance**

Our investigation commenced by analyzing the solvent concentration effect on IL/solvent electrolytes, employing [Bmim][PF$_6$] solvated in ACN with different IL-solvent ratios, to discern the optimal ratio. Simulation and experiment both show a volcano correlation between the electrolyte conductivity and the ACN concentration (**Supplementary Table 1** and **Supplementary Figure 1**). The conductivity increase is owing to the dissociation of ion pairing and the weakened interaction energy between cation and anion, leading to a faster ion diffusion (**Supplementary Figure 2**), while the decrease is ascribed to the domination of lessening IL.[13] Upon reaching an IL-solvent ratio of 0.107, the conductivity is increased by 13.1 times compared with the bulk value, peaking at 4.8 S m$^{-1}$. Meanwhile, the experiment showed the same optimal concentration, and the consistency between MD simulations and experimental measurements in **Supplementary Figure 1** pinpointed the optimized concentration for [Bmim][PF$_6$]/ACN electrolyte, hence, opted for investigating the solvent effect on c-MOF supercapacitors.

MD simulations with the constant potential method[14] were subsequently conducted for c-MOF supercapacitors with electrolytes of pure [Bmim][PF$_6$] and [Bmim][PF$_6$]/ACN electrolytes (**Figure 1a-b** and **Supplementary Table 2**). The charge storage of these supercapacitors is characterized by voltage-dependent capacitance and energy storage density (**Figure 1c-d**). The c-MOF with pure IL exhibited a U-shape of the capacitance-potential curve within a potential range of -1.5 V to 1.5 V, providing a predicted gravimetric capacitance of approximately 36 to 49 F g$^{-1}$. Remarkably, the introduction of ACN into the IL renders a bell-shaped curve, substantially leading to a one-fold increase in the capacitance, up to approximately 103 F g$^{-1}$ (**Figure 1c**). Consequently, there was a substantial improvement in gravimetric energy density as well (**Figure 1d**).

The charging dynamics were evaluated by examining the time evolution of accumulated charge on the MOF electrode under different applied voltages (**Figure 1e-f**). A complete charging process, defined as the time when the electrode charge reaches 95% of its maximum capacity[15], necessitated a minimum of 33 ns and resulted in an accumulated charge of around 40 C g$^{-1}$ at a cell voltage of 2 V in pure IL supercapacitor. In contrast, the IL/solvent-based electrolyte accelerated the charging process dramatically, completing it at around 4.5 ns (**Figure 1g**) and accumulating near-doubled charges on the MOF electrode. These findings indicate that the incorporation of organic solvent into ILs can offer a remarkable 104% enhancement in capacity and an about 6-fold increase in charging dynamics simultaneously. These enhancements exceed those reported in the literature for supercapacitors with porous carbon electrodes in ILs by adding solvent (**Supplementary Note 2** and **Supplementary Table 3**).[9-10, 16] These phenomena suggest that the regular and monodisperse pores of c-MOFs may play a vital role in the energy storage.

Moreover, the influence of solvent concentration on energy storage performance was explored (**Supplementary Figure 3**), for different IL-solvent ratios between 0.05 and 0.2 which could attain high conductivity (**Supplementary Figure 1**). These studied concentrations had a minor effect on charge storage capacity at the same applied voltage, but heavily affected the charging process. Notably, the electrolyte with optimal bulk ionic conductivity (IL-solvent ratio of 0.107) also exhibited the fastest charging dynamics.

**2.2 Quantitative comparison with experiment**

To validate the MD-obtained capacitance and charging performance, we carried out electrochemical measurements of c-MOF supercapacitors. Initially, the Ni$_3$(HITP)$_2$ crystallite powder was synthesized through a solvothermal reaction between the nickel salt and the organic linker 2,3,6,7,10,11-hexaaminotriphenylene hexahydrochloride (HATP·6HCl) at 65 °C (refer to **Methods** for details). The powder X-ray diffraction (XRD) analysis revealed a crystalline structure exhibiting several distinct peaks (**Figure 2a**), in line with simulated XRD and the previously reported Ni$_3$(HITP)$_2$ structure[3a, 17], indicating a well-defined crystal of Ni$_3$(HITP)$_2$ (**Supplementary Note 3** and **Supplementary Figure 4**). Intriguingly, the N$_2$ adsorption isotherm analysis elucidated an exceptional Brunauer-Emmett-Teller SSA of 1023 m$^2$ g$^{-1}$ (**Figure 2b**), almost approaching the theoretical value (1153 m$^2$ g$^{-1}$), larger than the highest value for Ni$_3$(HITP)$_2$ reported in the literature (884.7 m$^2$ g$^{-1}$)[18] (**Figure 2c** and **Supplementary Table 4**). This enhancement could be attributed to the preheating for crystallization, coupled with impurity removal through the solvent exchange during the synthesis process.

Two-electrode cells were meticulously assembled using near-ideal Ni$_3$(HITP)$_2$ samples as sole electrode materials to dissect their energy storage performance in both pure [Bmim][PF$_6$] and [Bmim][PF$_6$]/ACN electrolytes (**Methods** and **Supplementary Figures 4-5**). CV measurements at a scan rate of 10 mV s$^{-1}$ along with GCD of the symmetrical cell show nearly rectangular and triangular traces, respectively, displaying characteristics of capacitive behavior (**Figure 2d-e**). The specific gravimetric capacitance of a single electrode with pure IL is about 52 F g$^{-1}$ at a cell voltage of 1 V, closely aligning with the MD-derived capacitance of 41 F g$^{-1}$ at the same voltage. A quantitative agreement in the gravimetric capacitance was also observed for the c-MOF with IL/ACN electrolyte (MD-predicted 103 F g$^{-1}$ vs. experimental 112 F g$^{-1}$). Moreover, the IL/ACN supercapacitor exhibited exceptional cycling stability, retaining over 90% of its initial capacitance after 20,000 cycles at 5 A g$^{-1}$ (**Figure 2f**), showcasing superior stability compared to prior c-MOF-based supercapacitors.[3b-d]

To make an in-depth comparison of capacitive performance from experiment and MD simulation, we efficaciously developed multiscale circuit models, addressing the canonical transmission line model in both CV and GCD charging modes through the superposition principle and dimensionless processing (Details refer to **Methods**, **Supplementary Note 4** and **Supplementary Figures 6-8**). Specifically, the current ($I$) passing through the c-MOF electrode ($L$) for the CV measurement is precisely ascertained, as:

$$I = \frac{4U_0\omega}{RL} \sum_{n=1}^{\infty} \int_0^t SquareWave(\omega\tau) e^{-\frac{a^2(1-2n)^2\pi^2(t-\tau)}{4L^2}} d\tau \qquad (1)$$

where $a = \sqrt{1/(RC)}$, and $U_0$ represents the peak voltage; $R$ and $C$ signify the ion diffusion resistance and the capacitance per unit length, respectively; $n$ is the order unity of branches.[19]

The frequency of the charging period $T$ is denoted by $\omega$. $SquareWave$ represents a non-sinusoidal periodic waveform. Similarly, the GCD curve can be expressed as (**Supplementary Note 4**):

$$U = \frac{I_0 R}{3L}(L^2 + 3a^3 t)\theta(t - t_0) - \sum_{n=1}^{\infty} \frac{2I_0 RL}{n^2 \pi^2} e^{-\frac{a^2 \pi^2 n^2 (t-t_0)}{L^2}} \theta(t - t_0) \quad (2)$$

where $I_0$ is the applied constant current, and $\theta(t - t_0)$ is the unit step function about $t$.

Different from the extended transmission line model employed for porous carbon electrodes, which necessitates fitting parameters to account for the intricate topology of porous carbon[19-20], our models expressed in **Equations 1-2** demonstrate efficacy without additional fitting parameters, relying solely on the intrinsic characteristics of MOFs and electrolytes, specifically the MD-derived capacitance and ionic conductivity within the nanopore. Remarkably, these multiscale models consistently yield good alignments between the predicted and experimental CV and GCD curves (**Figure 2d-e**), highlighting that c-MOFs with their singular pore structure make them an excellent platform for modeling-experiment comparison through multiscale models.

## 2.3 Origin of solvent-enhanced charge storage

Understanding the solvent impact on energy storage was conducted at an atomic scale. The ion distribution within a quasi-1D nanopore was methodically examined along axial and planar cross-sections (**Figure 3**, **Supplementary Note 5,** and **Supplementary Figure 9**). In the pure IL supercapacitor, a substantial influx of both cations and anions into the MOF pore under 0 V indicated favorable IL wettability, in line with nuclear magnetic resonance (NMR) measurements for porous carbon electrodes.[21] From a 1D view along the pore side, cations and anions are arranged in interleaved layers, forming cation-anion layers in an interleaved manner along the pore axis (**Figure 3a**). A 2D ion distribution map in the planar cross-section depicted both cations and anions forming a layer adhered to the pore surface, presenting a hexagonal pattern, with an ion-wire along the pore axis (**Supplementary Figure 10**). Conversely, the in-pore ion populations dropped as ACN molecules predominantly occupied, resulting in an enhancement of the ionophobicity of MOF pores due to steric hindrance effects, consistent with the NMR measurement for the YP-50F electrode.[21] This solvent-induced ionophobicity of nanopore could facilitate the capacitance increase.[22] A similar wave-like ion distribution along the pore axis was observed for IL/ACN electrolyte (**Figure 3b**). Meanwhile, it was demonstrated that ACN molecules were closer to the surface compared with ions (**Figure 3c** and **Supplementary Figure 11**).

Under electrode polarizations, a disparity in ion distributions within the nanopore was evident. In the pure IL supercapacitor, co-ions persisted in the surface region with a slight displacement from the MOF surface (**Figure 3c**). In contrast, co-ions in IL/ACN electrolyte were completely expelled from the surface region. Instead, the surface region was dominated by counterions and ACN molecules, exhibiting a distinctive hexagonal distribution pattern with a wave-like motif along the planar cross-section (**Supplementary Figure 11**). Quantitatively, the deviation in the accumulated number density of counterions and co-ions remains marginal for pure IL; while this discrepancy experiences a pronounced increase under the same voltage with the addition of ACN (**Figure 3d** and **Supplementary Figure 12**), suggesting that more net charge could be stored in MOF pores, leading to a higher energy storage capacity featured as about onefold increase in capacitance (**Figure 1c**).

The solvent effect was further elucidated based on the local structure (**Figure 3e-f**, and **Supplementary Figure 13**). In pure IL, the radial distribution functions of cations and anions within the MOF electrode pore exhibit a discernible peak with minimal variation in the magnitude under both negative and positive polarizations (**Figure 3e**). This implies that the voltage struggles to separate cation and anion, resulting in a slight change in the coordination number of ions in the pore (**Figure 3g**). Conversely, in the presence of ACN, although a similar peak location is observed, there is a significantly reduced magnitude across the entire applied voltage region (**Figure 3f**). Quantitatively, the in-pore coordination number of cation and anion decreased from 4.8 to lower than 1, when adding ACN (**Figure 3g**). These phenomena indicate that ACN plays an ion pair disruptor, resulting in a noticeable increase in the percentage of free counterions within the MOF pore, thereby an increase in the net charge stored in the electrode (**Figure 3h** and **Supplementary Figure 14**). The unpairing effect of ACN could be understood from an energetic perspective where the separation energy between cations and anions within the pore for pure IL system (1.6 eV) was significantly decreased to around 0.7 eV upon the addition of ACN (**Figure 3i**).

Briefly, ACN molecules act as 'ionophobic agents' effectively controlling pore ionophobicity,[22] by predominantly occupying the pore under the entire polarized region. Additionally, they disrupt the coupling of cations and anions, facilitating the effective cation-anion separation and then increasing the net charge stored in the electrode under all applied voltages.[3c] These effects synergistically enhance energy storage capacity.

**2.4 Mechanism of solvent-accelerated charging dynamics**

The solvent effect on the charging process under various potentials was firstly quantified with the charging mechanism parameter ($X$):[21b]

$$X = \frac{N - N_0}{(N^{counter} - N^{co}) - (N_0^{counter} - N_0^{co})} \tag{3}$$

where $N$ and $N_0$ are the total numbers of ions inside a pore at a working potential and 0 V, respectively; $N^{counter}$ and $N_0^{counter}$ refer to counterions, and $N^{co}$ and $N_0^{co}$ denote the co-ions in MOF pores, respectively. The $X$ values for the pure IL supercapacitor, depicted in **Figure 4a**, indicate that the charge storage involves a combination of ion exchange and counterion adsorption, with $X$ approaching 0.19 in the positively charged electrode; while in the moderately negative electrode, charging primarily proceeds through ion exchange, accompanied by a small amount of co-ion desorption ($X < 0$), suggesting the pronounced involvement of [PF$_6$]$^-$ anions in the charge storage process. This aligns with the charging process observed with the same IL using carbide-derived carbon (CDC) electrodes.[23] However, under potentials lower than -1 V, the charging process transitions to counterion adsorption. Upon the addition of ACN, a similar charging process (*i.e.*, a combination of ion exchange and counterion adsorption) was found under the positive polarization with a higher value of $X$, suggesting that anions become more 'active' with adding ACN. Under the negative polarization, a notable change was observed: the charging still involves an ion exchange with slight counterion adsorption, signifying that [Bmim]$^+$ cations are more 'active'. This charging pattern differs from that observed in porous carbon supercapacitors with the same electrolyte ([Bmim][PF$_6$]/ACN) which demonstrated that anions are more 'active' and dominate the charge storage process[10b, 21b], emphasizing the substantial impact of electrode topology and organic solvent on charging dynamics.

We then dissect the spatiotemporal evolution of ions within the pore to gain comprehensive insights into the anatomy of ion transport (**Figure 4b-c** and **Supplementary Figures 15-16**). In terms of pure IL supercapacitor, the number density of counterions and co-ions remains relatively constant in the central region under negative polarization; while the ion-exchange phenomena between inside/outside the c-MOF pore manifest at the surface region, characterized by a marginal decrease in co-ions and a concomitant increase in counterions. More importantly, the 2D map, illustrating the evolution of ion transport paths over time (**Figure 4d-e**) and the ion trajectories during the charging process, discloses an overlapped trajectory of co-ions and counterions (**Supplementary Figure 17**), resulting in a discernible counterion and co-ion collisions within the MOF pore, akin to a 'traffic jam' of ions. Meanwhile, the ion exchange between the surface and central regions is minimal. Analogously, the evolution of ions under positive electrodes mirrors this complicated pattern (**Supplementary Figures 18-19**). Briefly, motion paths for ions during the charging process involve counterions inserting into the surface region, alongside co-ion exclusion within the same region, leading to ion collisions.

In sharp contrast, the solvent prompts ion participation from both surface and central regions during the charging process. Specifically, counterions experience an increase in the surface region alongside a reduction in the central region, with a reverse trend observed for co-ions (**Figure 4b-c**). These changes in density seem to cause the ion 'traffic jam' similar to the IL-based system (**Figure 4d-e**); however, a meticulous examination of ion transport pathways reveals a nuanced pattern (**Figure 4f-g** and **Supplementary Figure 17**): initially entangled co-ions in the surface region undergo a surface-to-central ion transition, progressively migrating to the central region (positioned between 0.16 to 0.4 nm away from the axis of a MOF pore), resulting in a depletion of co-ions within approximately 5 ns; simultaneously, these co-ions, along with those initially located in the central region, migrate outward from the pore through this distinct channel (**Supplementary Figure 19**). Nevertheless, in term of counterions, they could ingress into the MOF nanopore through the co-ion-depleted surface region pathway. Furthermore, we observed a counterintuitive phenomenon that a minor fraction of counterions in the central region (predominantly located at 0 to 0.20 nm from the pore axis) migrate out the pore, which is attributed to the attraction by the co-ions. Briefly, after the addition of solvent, two separated motion paths emerge for counterions and co-ions: co-ions leave the surface region to the central region and then get out of the pore along the central region; whereas counterions enter the pore through the co-ion-depleted surface region, avoiding ion 'traffic jam'.

Why could solvent-induced separated ion motion paths accelerate the charging dynamics? The quantification of ion migration was undertaken through a rigorous analysis of the survival probability function $F(t)$ to understand how ions enter/leave the pore and transfer between surface and central areas[7d, 24]:

$$F(t) = \frac{N(t_0, t_0 + t)}{N(t_0)} \qquad (4)$$

where $N(t_0)$ represents the number of ions in a certain region at $t_0$, and $N(t_0, t_0 + t)$ corresponds to the number of ions persisting in such region during $t_0 + t$. Consequently, the residence time ($\tau_r$) is calculated by $\tau_r = \int_0^\infty F(t)dt$.[7d, 24] As exhibited in **Figure 4h**, counterion survival probabilities in surface and central regions under -1 V indicate a long-living state in their initial region for the pure IL. Conversely, with ACN addition, a rapid attenuation is observed both in the surface and central regions. Notably, the average lifetime for a counterion into the surface region was reduced from about 6911 to 1087 ps in the presence of ACN. Meanwhile, we observed that the temporal dynamics of co-ion exclusion, leaving from the surface to the central region and subsequently exiting the pore, are estimated to occur within approximately 61 and 215 ps, respectively. Therefore, the migration of counterions into the surface region emerges as the rate-determining step in the charging process. It is noteworthy that an expected increase in charging dynamics was not found in hierarchical porous carbons in IL [Emim][BF$_4$] with adding ACN;[25] even a deceleration in ion diffusion was observed upon a certain ACN addition for CDC electrode supercapacitors.[9b] These differences might be attributed to the intricate topology of porous electrodes.[9b, 25] This underscores the potential of utilizing c-MOFs, characterized by regular pores and separated ion transport highways, for advancing electrode materials.

Upon further examination of the arrangement of cations in the surface region, it was observed that the added ACN molecules alter the ion arrangement significantly. Specifically, the ACN dipole screened the surface charge of the electrode, leading to a near disappearance of the arrangement of cations perpendicular to the axial direction, instead transitioning towards a parallel orientation along the axis direction (**Supplementary Figures 20-22**). This regulated arrangement promotes the movement of counterions in the same direction. Moreover, the solvent effect on the energy barrier for cation insertion into the c-MOF pore was then estimated via the mean force potential (PMF) methodology.[26] As depicted in **Figure 4i**, it was observed that cations encounter an initial energy barrier of 70 kJ mol$^{-1}$ for ingress into the pore under 0 V. However, the energy barrier is enormously reduced to nearly zero with the introduction of ACN, characterized by a nearly flat free energy curve, thereby accelerating the charging process.

Schematically illustrated in **Figure 4j**, the charging process predominantly involves an ion exchange in the surface region for pure IL supercapacitor, leading to a pronounced ion 'traffic jam'. Nevertheless, the addition of ACN solvent serves as an adept traffic controller of ions, meticulously modulating the ion transport within the c-MOF electrode through the strategic promotion of surface-central ion exchange and alleviating intricate ion transport bottlenecks of co-ions and counterions. That is, co-ions leave the surface region to the central region and then exit the pore along this channel; whereas counterions ingress the pore through the co-ion-depleted surface region, thereby avoiding ions 'traffic jam', which is absent in porous carbon electrode (*e.g.*, CDCs, hierarchical carbons, *etc.*)[9b, 25]. Meanwhile, ACN molecules alter the arrangement of counterion orientation. Therefore, added solvent dramatically reduces the energy barrier, concurrently facilitating ion insertion into the pore, notably diminishing the time for counterion getting into the surface region, which is the rate-determining step in the charging process (**Figure 4j**). These factors synergistically amplify the charging speed of c-MOF supercapacitors.

## 3. Conclusion

We systematically explored the impact of incorporating the organic solvent into pure IL on the energy storage of c-MOF supercapacitors. Employing constant potential MD simulations, we revealed a striking enhancement in both charge storage capacity and charging dynamics, receiving a >100% increase in capacitance and ~6 times improvement in charging speed after the addition of ACN with the optimal IL-solvent ratio of 0.107. To validate the modeling results, we successfully synthesized near-ideal c-MOF $Ni_3(HITP)_2$ with the largest SSA among all the literature and then fabricated them as supercapacitors. Quantitative alignment was achieved by theoretical models developed for linking microscopic modeling to macroscopic experimental energy storage performance, without any fitting parameters, underscoring the unique pore structure of c-MOFs as an excellent platform for theoretical research and multiscale amplification.

Essentially, the added ACN was found to act as 'ionophobic agent', effectively modulating pore ionophobicity by predominantly occupying the pore. Meanwhile, the microstructural analysis of IL and solvent in c-MOF pores, considering axial and in-plane distribution, revealed that ACN molecules weakened the interaction between cations and anions, leading to the decoupling of counterion and co-ion associations. This ionophobicity modulation, along with desolvation effects, synergistically facilitates an effective increase in charge storage capacity.

For charging dynamics, the solvent functions as a skilled 'traffic police' of ions within the c-MOF electrode to properly adjust the flow of ions, strategically facilitating ion exchange between the surface and central regions and easing complex ion transport bottlenecks for both counterions and co-ions. In particular, co-ions transport from the surface region to central region and then leave the pore via this pathway; whereas counterions enter the pore through the co-ion-depleted surface region, effectively avoiding ion 'traffic jam'; the ion transport within the surface region is the rate-determining step in the charging process; ACN molecules modify the arrangement of counterion orientation and lower the energy barrier. These factors amplify charging speed within the pore.

Our findings have provided an in-depth understanding of the solvent effect on the energy storage mechanism of c-MOF supercapacitors and established a bridge to link the microscopic simulations and macroscopic electrochemical measurements. These insights could not only lay the groundwork for designing porous supercapacitors with best both high energy and power performance, but also benefit other applications such as battery[27], IL gating[28], and electrowetting[29].

## 4. Methods

*Molecular dynamics simulations:* As shown in **Figure 1**, the MD simulation system consists of two identical and symmetric conductive MOFs immersed in [Bmim][PF$_6$] or [Bmim][PF$_6$]/ACN electrolytes. The atomistic structure of Ni$_3$(HITP)$_2$ was taken from our previous work.[5a] Lennard-Jones force field parameters for MOF atoms are derived from the generic universal force field.[30] As for the electrolyte, coarse-grained force fields are adopted for [Bmim][PF$_6$] and ACN,[31] which has been proven to correctly simulate the solution while conserving the computational resource significantly.[10b, 32] Both the positive and negative electrodes exhibit symmetric structures, each composed of a stack of 18 conductive MOF layers. The distance between the centers of the two electrodes is fixed at 20 nm, which ensures non-interaction between them during the MD simulation process. In the supercapacitor based on pure IL, the electrolyte contains 1540 pairs of ions, whereas in the supercapacitor with IL/ACN, there are 486 pairs of ions and 4532 ACN molecules, corresponding to the optimal IL-solvent ratio of 0.107 (**Supplementary Table 2**).

Simulations were conducted using the customized MD software GROMACS.[14, 33] Specifically, the constant-potential method (CPM)[5a, 14] was employed to allow the fluctuations of the charges on electrode atoms to ensure an adequate description of the electrode polarization effects in the presence of electrolytes. The V-rescale thermostat was utilized to couple the temperature to 298 K in the NVT ensemble.[34] The cutoff radius for short-range electrostatic and van der Waals interactions was set at 1.2 nm, while long-range electrostatic interactions were computed using the particle mesh Ewald (PME) algorithm.[35] The PME k-space grid spacing was set to 0.1 nm. The simulation system was initially simulated at a temperature of 1000 K for 20 ns to achieve preliminary equilibrium. Subsequently, a simulated annealing algorithm was applied to couple the temperature to 298 K over a 10 ns simulation period, followed by maintaining this temperature for 50 ns to generate a preliminary equilibrium configuration under null electrode potential. To obtain microstructure and capacitance, a simulation was performed for 50 ns to ensure reaching equilibrium under applied voltages ranging from 0 to 5 V, and then another 20 ns production in the equilibrium state was run for analysis. To investigate the charging dynamics, five independent runs were performed to certify the accuracy of the simulation results.

*Materials for experiments:* All starting materials were procured from commercial sources and employed without additional purification unless explicitly stated. Specially, nickel(II) acetate tetrahydrate (Ni(OAc)$_2$·4H$_2$O), NaOAc and ACN were purchased from Aladdin; 2,3,6,7,10,11-hexaaminotriphenylene hexahydrochloride (HATP·6HCl) was purchased from Yanshen Technology Co., Ltd.; N,N-dimethylformamide (DMF), N,N-dimethylacetamide (DMA) and methanol were purchased from Sinopharm Chemical Reagent Co., Ltd.. IL [Bmim][PF$_6$] (Lanzhou Yulu Fine Chemical Co., Ltd) was purified via the Schlenk line at 85 °C for 48 h. ACN was purified by P$_2$O$_5$ and subsequently collected via distillation and stored in a desiccator until required.

*Conductivity measurements of electrolytes:* Conductivities of electrolytes with different IL-solvent ratios were measured using a conductivity meter from METTLER TOLEDO, model S230 SevenCompact, operating at room temperature conditions. The instrument precision is specified at 0.5% for individual measurements. Prior to measurement, IL and solvent were meticulously mixed within a sealed glass vessel, ensuring homogeneity, and data acquisition was commenced only upon attaining a stable equilibrium state.

*Synthesis of c-MOFs:* A total of 120 mL of a stock solution (2.33 mmol L$^{-1}$) of Ni(OAc)$_2$·4H$_2$O in DMF/DMA (v/v = 1:1) was preheated to 65 °C in a 500 mL beaker. Subsequently, 80 mL of fresh NaOAc aqueous solution (2 mol L$^{-1}$) was added under agitation. The resulting solution was then heated to 65 °C before incorporating a solution containing 100 mg of HATP·6HCl in 30 mL of water. This mixture underwent high-speed stirring for 4 hours at 65 °C. Following this, this suspension was allowed to cool to room temperature. The supernatant solution was removed, and the resultant black precipitate was subjected to centrifugation, followed by solvent exchange with water (180 mL×2) and methanol (180 mL×2). In a solvent-exchanged process, a mixture of solvent and MOF powder was agitated for at least 40 minutes at room temperature, and the solid was then filtered off. Finally, the residue was subjected to vacuum conditions.

*Characterization of c-MOF powder:* Nitrogen adsorption/desorption isotherms were measured with a Quantachrome Autosorb IQ system at 77 K. Before the gas adsorption/desorption measurement, as-synthesized MOFs (~70 mg) samples were activated by drying under vacuum for 1 day at 90 °C. Afterward, liquid nitrogen baths (77 K) were used to measure nitrogen adsorption/desorption isotherms. The free space correction and measurement were conducted using ultrahigh-purity grade (99.999% purity) nitrogen and helium, oil-free valves, and gas regulators. Powder X-ray direction (PXRD) patterns were carried out with a PANalytical Empyrean X-ray Powder diffractometer equipped with a Cu-sealed tube ($\lambda = 1.544426$ Å) at 40 kV and 40 mA, and samples were prepared on a zero-background silicon crystal plate. Scanning electron microscopy (SEM) was conducted on a ZEISS GeminiSEM 300 with an InLens detector at an operating voltage of 3 kV. High-resolution transmission electron microscopy images were obtained with a FEI Tecnai G2 20, operated at an accelerating voltage of 200 kV equipped with a Gatan K2 in-situ direct electron detector. Samples were drop-cast onto Cu TEM grids from powder dispersed in methanol.

*Electrochemical measurements of c-MOF supercapacitors:* A two-electrode cell was fabricated within an Argonne-filled glovebox to assess the capacitive performance of c-MOFs. The cell is assumed to be symmetric, since the difference of two electrodes was controlled within 5%. All electrochemical measurements were carried out using a Biologic VMP-3e potentiostat. A series of scan rates (50, 20, 10, 5, 1 mV s$^{-1}$) and current densities (5, 2, 1, 0.5, 0.1 A g$^{-1}$) were set to obtain the capacitive performance in CV and GCD measurements, respectively. Through this work, all c-MOF electrodes were fabricated without any binders or conductive additives. The electrode was made of MOF pellets in 7 mm diameter and 105 µm thickness (mass loading ~6 mg cm$^{-2}$).

We calculated the gravimetric capacitance ($C_g$) of one single electrode from CV curves, according to:

$$C_g = 4\frac{\int_0^{V_0/v} I\, dt}{V_0 m} \tag{5}$$

where $I$ and $v$ are the discharge current and the scan rate, respectively, $V_0$ is the operating voltage, and $m$ is the total active material mass of both electrodes in the symmetric cell.

$C_g$ can be compared with MD-obtained integral capacitance ($C_g^{MD}$). $C_g^{MD}$ can be obtained directly from the MD simulation that is performed under $V_0$, the potential drop between the two electrodes, as $C_g^{MD} = 4\frac{Q}{V_0 m}$, in which $Q$ is the total charge on one electrode.

## Supporting Information

Supplementary information is available from the Wiley Online Library or from the author.

## Acknowledgements

M.C., T.Z.W., and L.N. contributed equally to this work. Authors in HUST acknowledge the funding support from the National Natural Science Foundation of China (T2325012, 52106090, 52161135104) and the Program for HUST Academic Frontier Youth Team. A.A.K. thanks the grant from the Engineering and Physical Sciences Research Council (EP/L015579/1). M. C. also thanks the China Postdoctoral Science Foundation (2022T150228). Simulations in this work were accomplished in the Wuhan Supercomputing Center.

## Conflict of Interest

The authors declare no competing interests.

## Data Availability Statement

The data that support the findings of this study are available from the corresponding author upon reasonable request.

**Figures**

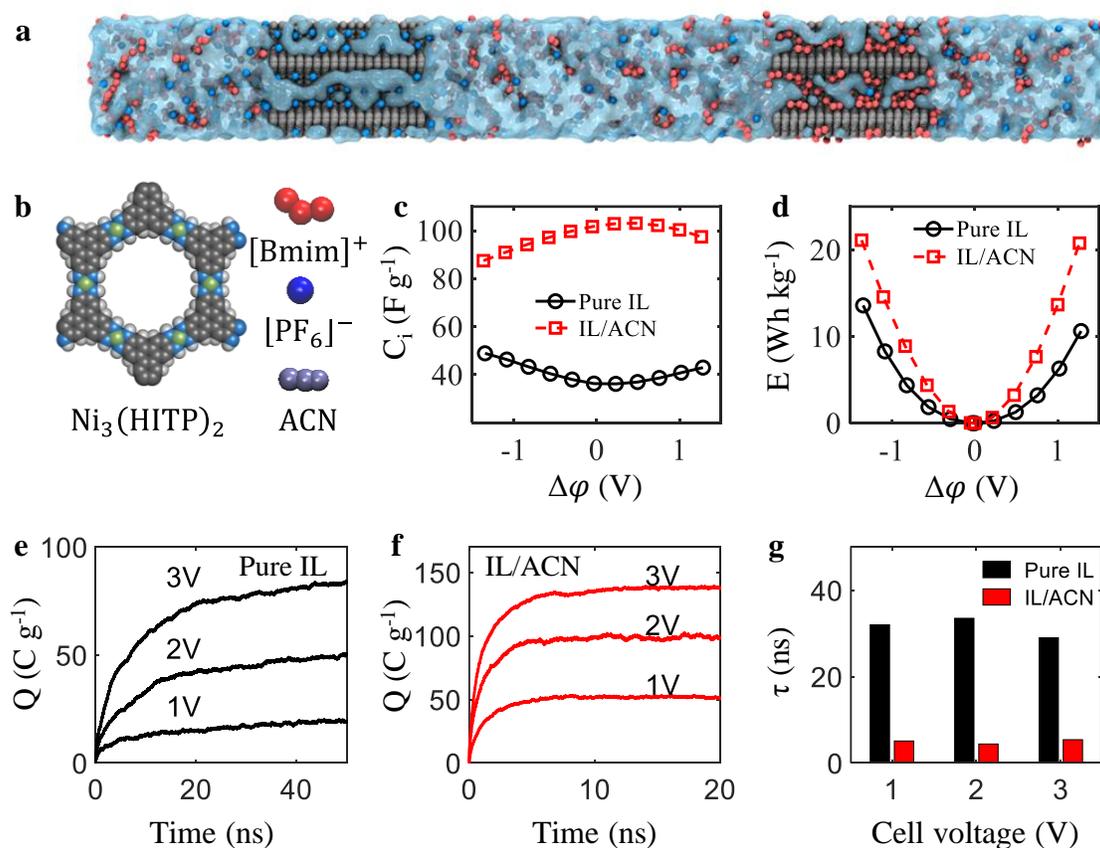

Figure 1. Solvent effects on energy storage performance of c-MOF supercapacitors. a, Snapshot of MD simulation systems. b, Atomic structure of electrodes and electrolytes. The red, blue, and violet spheres represent coarse-grained cations, anions, and ACN molecules, respectively. c,d, Gravimetric capacitance (c) and energy density (d) with different electrode voltages. e,f, Evolution of charges on MOF electrodes with pure [Bmim][PF$_6$] (e) and [Bmim][PF$_6$]/ACN electrolytes (f). g, Solvent effects on the charging time.

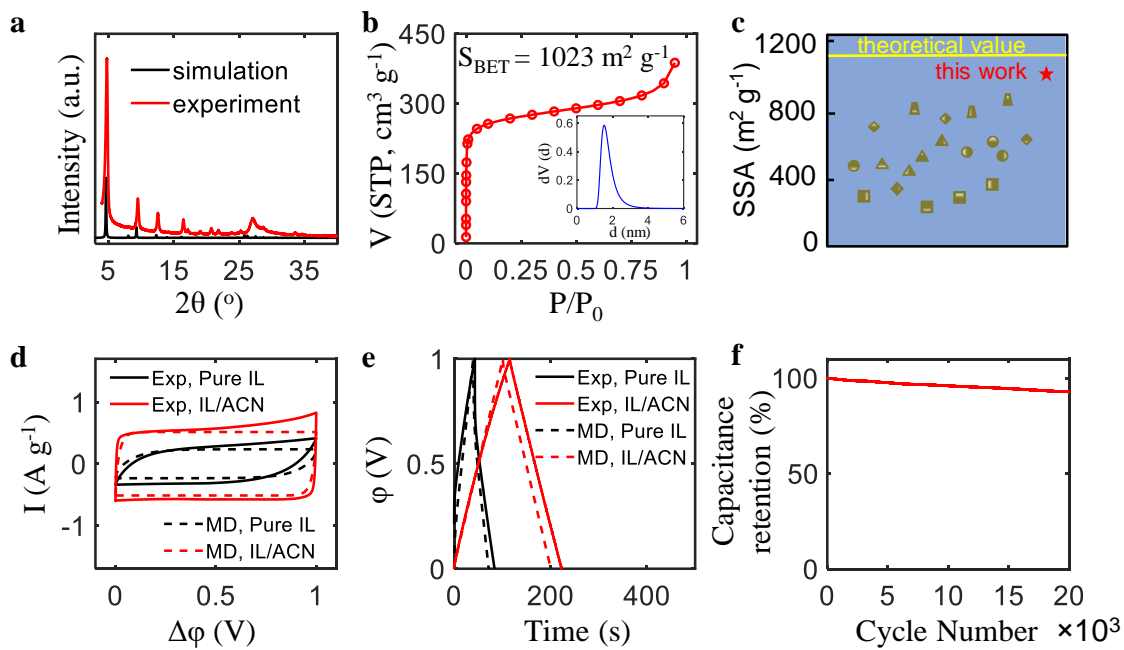

Figure 2. Measured capacitive performance of c-MOF supercapacitors. a, Experimental and simulated PXRD of $Ni_3(HITP)_2$. b, $N_2$ adsorption isotherms of $Ni_3(HITP)_2$ at 77 K. c, Comparison of SSA in this work and the reported $Ni_3(HITP)_2$ from the literature (see Supplementary Table 4). d-f, Comparison of c-MOF electrodes in a symmetrical supercapacitor for CV (d), and GCD (e) curves between experimental measurements and MD simulations. Scan rate of the CV test is 10 mV s$^{-1}$, and the current density of GCD measurement is 0.5 A g$^{-1}$. g, Capacitance retention under repeated cycling at a current density of 5 A g$^{-1}$ for 20,000 cycles.

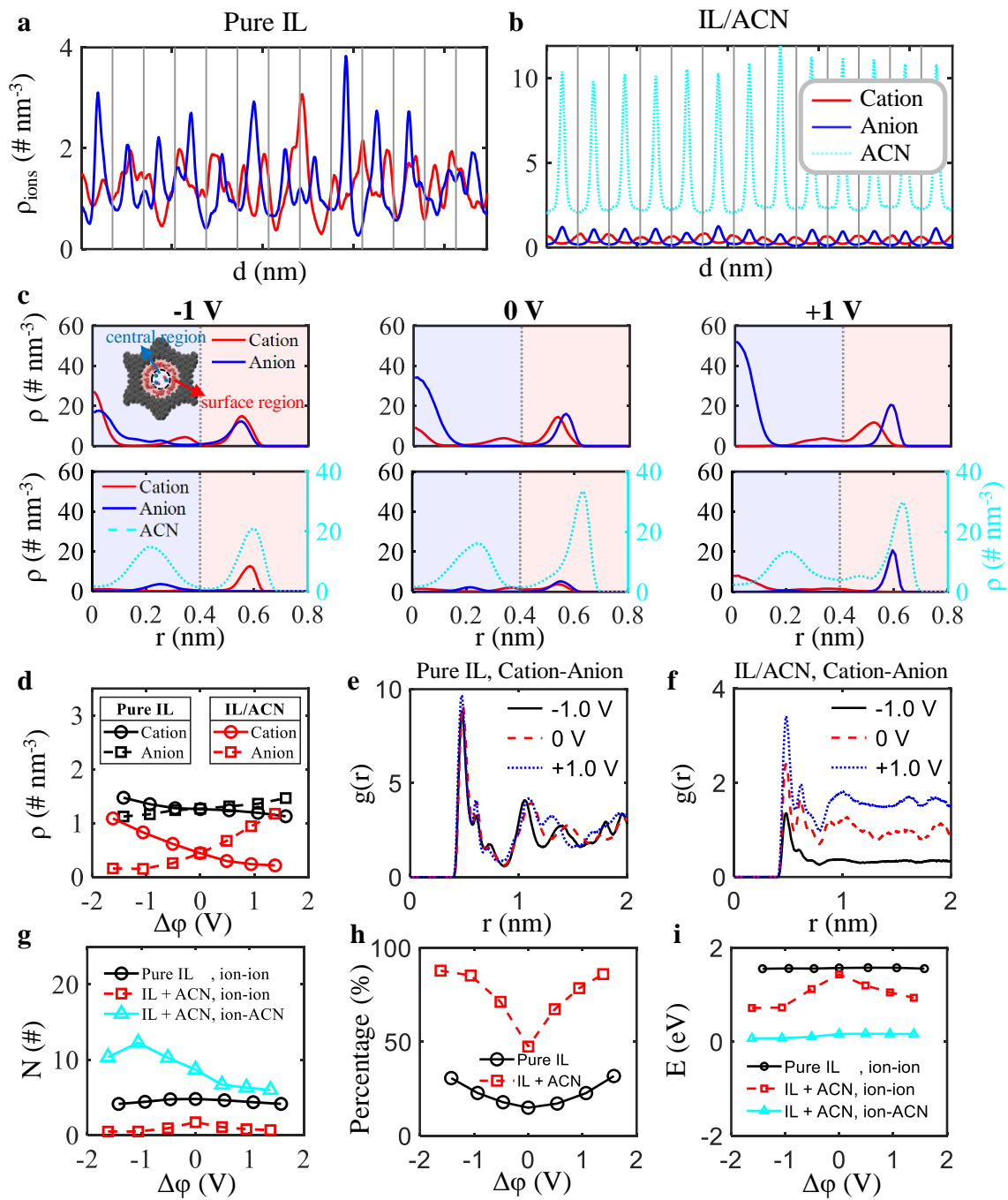

Figure 3. Ion structure inside c-MOF pores. a-b, Number density distributions of cations and anions along axial section inside MOF pores under 0 V for pure IL (a) and IL/ACN electrolytes (b). c, Radial number distributions of ions (or solvent molecules) inside the pore for pure IL and IL/ACN electrolytes. r = 0 means the axis of a MOF pore. d, Accumulative number density of ions inside c-MOF pores under various potentials. e-f, Radial distribution functions of ions inside MOF pores for pure IL (e) and IL/ACN (f) systems. g, Coordination number distributions of counterions and co-ions. h, Percentage of free counterions inside MOF pores. i, Separation energy between ions (or solvent molecules) inside positive charged c-MOF pores.

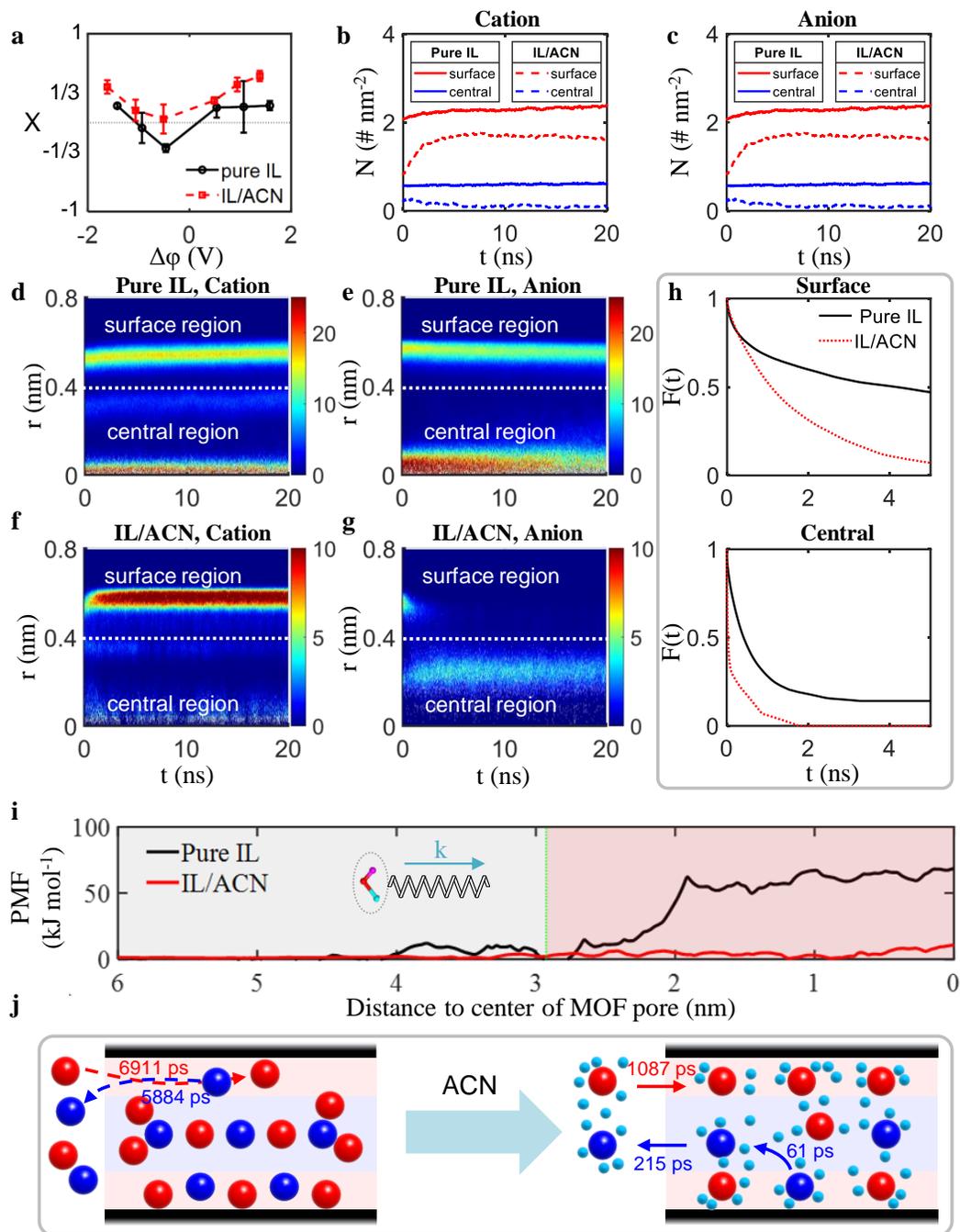

Figure 4. Mechanism of solvent-accelerated charging dynamics. a, Charging mechanism parameter, X, as a function of voltage. b-c, Evolution of cation (b) and anion (c) number density, N, inside negatively charged MOF pores during charging. d-e, Spatiotemporal evolution of the cation (d) and anion (e) inside the negative electrode for pure IL. f-g, Spatiotemporal evolution of the cation (f) and anion (g) inside the negatively charged pores for IL/ACN. h, Time-correlation function of ionic adsorption characteristic function for cations inside MOF pores at various electrode voltages. i, Potential of mean force (PMF) for moving a cation from the bulk electrolyte into the MOF pore. The green dotted line represents the outer surface of MOF electrode. j, Schematic of solvent effect on the ion migration. Red, blue, and wathet spheres represent cations, anions, and solvent molecules, respectively. The red and blue arrows indicate ion motion paths.

# TOC

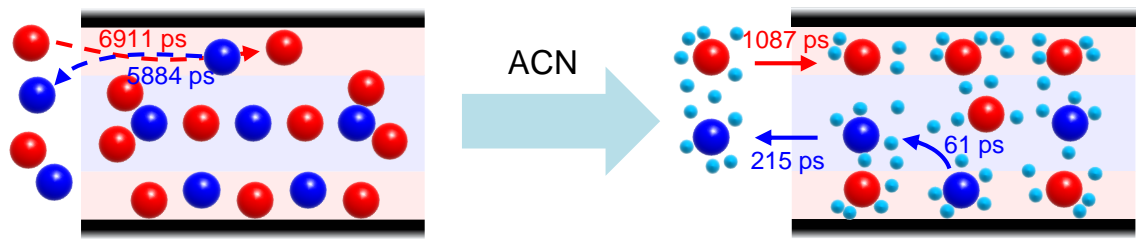